\documentstyle[epsfig]{mn}

\newif\ifAMStwofonts

\ifoldfss
  \newcommand{\rmn}[1] {{\rm #1}}

  \ifCUPmtlplainloaded \else
    \NewTextAlphabet{textbfit} {cmbxti10} {}
    \NewTextAlphabet{textbfss} {cmssbx10} {}
    \NewMathAlphabet{mathbfit} {cmbxti10} {} 
    \NewMathAlphabet{mathbfss} {cmssbx10} {} 
  \fi
  \ifAMStwofonts
    \ifCUPmtlplainloaded \else
      \NewSymbolFont{upmath} {eurm10}
      \NewSymbolFont{AMSa} {msam10}
      \NewMathSymbol{\upi}     {0}{upmath}{19}
      \NewMathSymbol{\umu}     {0}{upmath}{16}
      \NewMathSymbol{\upartial}{0}{upmath}{40}
      \NewMathSymbol{\leqslant}{3}{AMSa}{36}
      \NewMathSymbol{\geqslant}{3}{AMSa}{3E}

       \let\le=\leqslant
       
    \fi
  \fi
\fi 

\ifnfssone
  \newmathalphabet{\mathit}
  \addtoversion{normal}{\mathit}{cmr}{m}{it}
  \addtoversion{bold}{\mathit}{cmr}{bx}{it}
  \newcommand{\rmn}[1] {\mathrm{#1}}

  \newmathalphabet{\mathbfit} 
  \addtoversion{normal}{\mathbfit}{cmr}{bx}{it}
  \addtoversion{bold}{\mathbfit}{cmr}{bx}{it}
  \newmathalphabet{\mathbfss} 
  \addtoversion{normal}{\mathbfss}{cmss}{bx}{n}
  \addtoversion{bold}{\mathbfss}{cmss}{bx}{n}
  \ifAMStwofonts
    \ifCUPmtlplainloaded \else
      \UseAMStwoboldmath
      \makeatletter
      \new@mathgroup\upmath@group
      \define@mathgroup\mv@normal\upmath@group{eur}{m}{n}
      \define@mathgroup\mv@bold\upmath@group{eur}{b}{n}
      \edef\UPM{\hexnumber\upmath@group}
      \new@mathgroup\amsa@group
      \define@mathgroup\mv@normal\amsa@group{msa}{m}{n}
      \define@mathgroup\mv@bold\amsa@group{msa}{m}{n}
      \edef\AMSa{\hexnumber\amsa@group}
      \makeatother
      \mathchardef\upi="0\UPM19
      \mathchardef\umu="0\UPM16
      \mathchardef\upartial="0\UPM40
      \mathchardef\leqslant="3\AMSa36
      \mathchardef\geqslant="3\AMSa3E

       \let\le=\leqslant

    \fi
  \fi
\fi 

\ifnfsstwo
  \newcommand{\rmn}[1] {\mathrm{#1}}

  \DeclareMathAlphabet{\mathbfit}{OT1}{cmr}{bx}{it}
  \SetMathAlphabet\mathbfit{bold}{OT1}{cmr}{bx}{it}
  \DeclareMathAlphabet{\mathbfss}{OT1}{cmss}{bx}{n}
  \SetMathAlphabet\mathbfss{bold}{OT1}{cmss}{bx}{n}
  \ifAMStwofonts
    \ifCUPmtlplainloaded \else
      \DeclareSymbolFont{UPM}{U}{eur}{m}{n}
      \SetSymbolFont{UPM}{bold}{U}{eur}{b}{n}
      \DeclareSymbolFont{AMSa}{U}{msa}{m}{n}
      \DeclareMathSymbol{\upi}{0}{UPM}{"19}
      \DeclareMathSymbol{\umu}{0}{UPM}{"16}
      \DeclareMathSymbol{\upartial}{0}{UPM}{"40}
      \DeclareMathSymbol{\leqslant}{3}{AMSa}{"36}
      \DeclareMathSymbol{\geqslant}{3}{AMSa}{"3E}

       \let\le=\leqslant

    \fi
  \fi
\fi 

\ifCUPmtlplainloaded \else
  \ifAMStwofonts \else 
    \def\upi{\pi}
    \def\umu{\mu}
    \def\upartial{\partial}
  \fi
\fi

\title[Models of Rotating Neutron Stars: Remnants of AIC]{Models of Rapidly Rotating 
Neutron Stars: Remnants of Accretion Induced Collapse}
\author[Yuk Tung Liu and Lee Lindblom]
{Yuk Tung Liu and Lee Lindblom \\
Theoretical Astrophysics, California Institute of Technology, Pasadena, CA 91125}

\def\LaTeX{L\kern-.36em\raise.3ex\hbox{a}\kern-.15em
    T\kern-.1667em\lower.7ex\hbox{E}\kern-.125emX}

\begin{document}

\label{firstpage}

\maketitle

\begin{abstract}
Equilibrium models of differentially rotating nascent neutron stars 
are constructed, which represent the result of the accretion induced 
collapse of rapidly rotating white dwarfs. The models are built in a 
two-step procedure: (1) a rapidly rotating pre-collapse white dwarf 
model is constructed; 
(2) a stationary axisymmetric neutron star having the same total mass 
and angular momentum distribution as the white dwarf is constructed.
The resulting collapsed objects consist of a high density central core of 
size roughly 20~km, surrounded by a massive accretion torus extending over 
1000~km from the rotation axis. The ratio of the rotational kinetic energy 
to the gravitational potential energy of these neutron stars ranges from 
0.13 to 0.26, suggesting that some of these objects may have a non-axisymmetric 
dynamical instability that could emit a significant amount of gravitational 
radiation.
\end{abstract}


\section{INTRODUCTION}
\label{introduction}

The accretion induced collapse of a rapidly rotating white dwarf
can result in the formation of a rapidly and differentially
rotating compact object.
It has been suggested that such rapidly rotating objects could emit a substantial 
amount of gravitational 
radiation \cite{thorne95}, which might be observable by the gravitational wave 
observatories such as LIGO, VIRGO and GEO. It has been demonstrated that if 
the collapse is axisymmetric, the energy emitted by gravitational waves is 
rather small \cite{muller81,finn90,monchmeyer91,zwerger97}.
However, if the collapsed object rotates rapidly enough 
to develop a non-axisymmetric `bar' instability, the total energy released by 
gravitational waves could be $10^4$ times greater than the axisymmetric case
\cite{houser94,houser96,smith96,houser98}.

Rotational instabilities of rotating stars arise from non-axisymmetric 
perturbations of the form $e^{im\varphi}$, where $\varphi$ is the azimuthal 
angle. The $m=2$ mode is known as the {\em bar mode}, which is often the 
fastest growing unstable mode. There are two kinds of instabilities. A
{\em dynamical} instability is driven by hydrodynamics and gravity, and 
develops on a dynamical timescale, i.e.\ the time for sound waves to travel 
across the star. A {\em secular} instability is
driven by dissipative processes such as viscosity or gravitational radiation
reaction, and the growth time is determined by the dissipative timescale. 
These secular timescales are usually much longer than the dynamical timescale 
of the system. An interesting class of secular and dynamical instabilities 
only occur in rapidly rotating 
stars. One convenient measure of the rotation of a star is the parameter 
$\beta=T_{\rm rot}/|W|$, where $T_{\rm rot}$ is the rotational kinetic energy 
and $W$ is the gravitational potential 
energy. Dynamical and secular instabilities set in when $\beta$ exceeds the
critical values $\beta_d$ and $\beta_s$
respectively. It is well known that $\beta_d \approx 0.27$ and $\beta_s\approx 
0.14$ for uniformly rotating, constant density and incompressible stars, the
Maclaurin spheroids \cite{chandrasekhar69}. Numerous numerical 
simulations in Newtonian theory show that $\beta_d$ and $\beta_s$ have roughly 
these same values for differentially rotating polytropes with the same specific angular 
momentum distribution as the Maclaurin spheroids
\cite{tohline85,durisen86,williams88,houser94,smith96,houser96,pickett96,houser98,new99}. 
However, the critical values of $\beta$ are smaller for polytropes with some
other angular momentum distributions \cite{imamura95,pickett96,centrella00}. 
And general relativistic simulations also suggest that the 
critical values of $\beta$ are smaller than the classical Maclaurin spheroid 
values \cite{stergioulas98,shibata00,saijo00}.

Most of the stability analyses to date have been carried out on stars having 
simple {\em ad hoc} rotation laws. 
It is not clear whether these rotation laws are appropriate for the 
nascent neutron stars formed from the accretion induced collapse of rotating 
white dwarfs.

New-born neutron stars resulting from the core collapse of massive 
stars with realistic rotation laws were studied by M\"{o}nchmeyer, Janka and M\"{u}ller 
(M\"{o}nchmeyer \& M\"{u}ller 1988; Janka \& M\"{o}nchmeyer 1989ab), and 
Zwerger and M\"{u}ller \shortcite{zwerger97}. The study of 
M\"{o}nchmeyer et al.\ shows that the resulting neutron stars have $\beta<0.14$. 
Zwerger and M\"{u}ller carried out simulations of 78 models using simplified 
analytical equations of state (EOS). They found 
only one model having $\beta > 0.27$ near core bounce. However, $\beta$ remains 
larger than 0.27 for only about one millisecond, because the core re-expands 
after bounce and slows down. 
The pre-collapse core of that model is the most extreme one in their 
large sample: it is the most rapidly and most differentially rotating model, 
and it has the softest EOS. In 
addition, they found three models with $0.14<\beta<0.27$.
Rampp, M\"{u}ller and Ruffert \shortcite{rampp98} subsequently performed 3D 
simulations of three of these models. 
They found that the model with $\beta>0.27$ shows a non-linear growth of a
non-axisymmetric dynamical instability dominated by the bar mode ($m=2$). 
However, no instability is observed for the other two models during their 
simulated time interval of tens of milliseconds, suggesting that they are 
dynamically stable. 
Their analysis does not rule out the possibility that these models have 
non-axisymmetric secular instabilities, because the secular timescale 
is expected to range from hundreds of milliseconds to few minutes, much longer than 
their simulation time.

The aim of this paper is to improve Zwerger and M\"{u}ller's study by using 
realistic EOS for both the pre-collapse white dwarfs and the 
collapsed stars. For the pre-collapse white dwarfs, we use the EOS
of a zero-temperature degenerate electron gas with electrostatic corrections.
A hot, lepton rich protoneutron star is formed as a result of the collapse. 
This protoneutron star cools down to a cold neutron star in about 20~s (see 
e.g.\ Burrows \& Lattimer 1986), which is much longer than the dynamical timescle. 
So we adopt two EOS for the collapsed stars: one is suitable for 
protoneutron stars; the other is one of the standard 
cold neutron-star EOS. 

Instead of performing the complicated 
hydrodynamic simulations, however, we adopt a much simpler method. We assume 
(1) the collapsed stars are in rotational equilibrium with no meridional circulation, 
(2) any ejected material during the collapse carries a negligible amount of mass 
and angular momentum, and
(3) the neutron stars have the same specific angular momentum distributions 
as those of the pre-collapse white dwarfs. 
The justifications of these assumptions will be discussed in Section~\ref{col-mod}.
Our strategy is as follows. First we build the 
equilibrium pre-collapse rotating white dwarf models and calculate their 
specific angular momentum distributions. Then we construct the resulting collapsed
stars having the same masses, total angular momenta and 
specific angular momentum distributions as those of the pre-collapse white 
dwarfs. All computations in this paper are purely Newtonian. 
In the real situation, if a dynamical instability occurs, the star will never 
achieve equilibrium. However, our study here can still give a useful clue 
to the instability issue.

The paper is organized as follows. In the next section we present equilibrium 
models of pre-collapse, rapidly and rigidly rotating white dwarfs. In 
Section~\ref{col-mod}, we construct the equilibrium models corresponding to 
the collapse of these white dwarfs.
The stabilities of the collapsed objects are discussed in Section~\ref{stability}. 
Finally, we summarize our conclusions in Section~\ref{conclusion}.

\section{PRE-COLLAPSE WHITE DWARF MODELS}
\label{ini-mod}

\subsection{Collapse mechanism}

As mass is accreted onto a white dwarf, the matter in the white dwarf's interior
is compressed to higher densities. Compression releases 
gravitational energy and some of the energy goes into heat \cite{nomoto82}. 
If the accretion rate is high enough, the rate of heat generated by 
this {\em compressional heating} is greater than the cooling rate and the 
central temperature of the accreting white dwarf increases with time.

The inner core of a carbon-oxygen (C-O) white dwarf becomes unstable when the 
central density 
or temperature becomes sufficiently high to ignite explosive carbon burning. 
Carbon deflagration releases nuclear energy and causes the pressure to 
increase. However, electron capture behind the carbon deflagration front 
reduces the temperature and pressure and triggers collapse.
Such a white dwarf 
will either explode as a Type~Ia supernova or collapse to a neutron star.
Which path the white dwarf takes depends on the competition between 
the nuclear energy release and electron capture \cite{nomoto87}. 
If the density at which carbon ignites is higher than a critical density of about 
$9\times 10^9~\rmn{g}~\rmn{cm}^{-3}$ \cite{timmes92}, electron capture takes over and 
the white dwarf will collapse to a neutron star. However, if the ignition 
density is lower than the critical density, carbon deflagration will lead to a total 
disruption of the whole star, leaving no remnant at all. More recent calculations
by Bravo \& Garc\'{\i}a-Senz \shortcite{bravo99}, taking into account the 
Coulomb corrections to the EOS, suggests that this critical density
is somewhat lower: $6\times 10^9~\rmn{g}~\rmn{cm}^{-3}$. The density at 
which carbon ignites depends on the central temperature. The central temperature is 
determined by the balance between the compressional heating and the cooling and so 
strongly depends on the accretion rate and accretion time.
For zero-temperature C-O white dwarfs, carbon ignites at a density of about
$10^{10}~\rmn{g}~\rmn{cm}^{-3}$ \cite{salpeter69,ogata91}, which is higher
than the above critical density. If the accreting white dwarf can somehow 
maintain a low central temperature during the whole accretion process, 
carbon will ignite at a density higher than 
the critical density, and the white dwarf will collapse to a neutron star.
The fate of an accreting white dwarf as a function of the accretion rate and 
the white dwarf's initial mass is summarized in 
two diagrams in the paper of Nomoto \shortcite{nomoto87} (see also Nomoto \& 
Kondo 1991). Roughly speaking, low accretion rates 
($\dot{M}\la 10^{-8} M_{\odot}~\rmn{yr}^{-1}$) and high initial mass of the 
white dwarf ($M \ga 1.1\, M_{\odot}$), or very high accretion rates (near the 
Eddington limit) lead to collapse rather than explosion.

Under certain conditions, an accreting oxygen-neon-magnesium (O-Ne-Mg) white dwarf
can also collapse to a neutron star \cite{nomoto87,nomoto91}.
The collapse is triggered by the electron captures of $^{24}\rmn{Mg}$ and 
$^{20}\rmn{Ne}$ at a density of $4\times 10^9~\rmn{g}~\rmn{cm}^{-3}$. Electron 
captures not only soften the EOS and induce collapse, but also 
generate heat by $\gamma$-ray emission. When the star is collapsed to a central 
density of $10^{10}~\rmn{g}~\rmn{cm}^{-3}$, oxygen ignites \cite{nomoto91}. 
At such a high density, however, electron captures occur at a faster rate than 
the oxygen burning and the white dwarf collapses all the way to a neutron star.

In this section, we explore a range of possible pre-collapse white 
dwarf models. 
We assume that the white dwarfs are rigidly rotating. This is justified 
by the fact that the timescale for a magnetic field to suppress any 
differential rotation, $\tau_B$, is short compared with the
accretion timescale. For example, $\tau_B \sim 10^3~\rmn{years}$ if the 
massive white dwarf has a magnetic field $B\sim 100G$.
We construct three white dwarf models using the EOS of a zero-temperature 
degenerate 
electron gas with Coulomb corrections derived by Salpeter \shortcite{salpeter61}.
All three white dwarfs are rigidly rotating at the maximum possible angular 
velocities. {\em Model~I} represents a C-O white dwarf with a central density of 
$\rho_c=10^{10}~\rmn{g}~\rmn{cm}^{-3}$, the highest $\rho_c$ a C-O white 
dwarf can have before carbon ignition induces collapse. {\em Model~II} is 
also a C-O white dwarf but has a lower central density, $\rho_c=6\times 
10^9~\rmn{g}~\rmn{cm}^{-3}$. This is the lowest central density for which 
a white dwarf can still collapse to a neutron star after carbon ignition. 
{\em Model~III} is an O-Ne-Mg white dwarf with 
$\rho_c=4\times 10^9~\rmn{g}~\rmn{cm}^{-3}$, that is the density at which
electron captures occur and induce the collapse.
Since the densities are very high, the pressure 
is dominated by the ideal degenerate Fermi gas with electron fraction $Z/A=1/2$ 
that is suitable for both C-O and O-Ne-Mg white dwarfs. Coulomb corrections, 
which depend on the white dwarf composition through the atomic number $Z$,
contribute only a few per cent to the EOS at high densities, so the 
three white dwarfs are basically described by the same EOS. 

\subsection{Numerical method}

We treat the equilibrium rotating white dwarfs as rigidly rotating, 
axisymmetric, and having no internal motion other than the motion due to 
rotation. The Lichtenstein theorem (see Tassoul 1978) guarantees that 
a rigidly rotating star has reflection symmetry about the equatorial plane.
We also neglect viscosity and assume Newtonian gravity. Under these 
assumptions the equilibrium configuration is described by the static Euler 
equation:
\begin{equation}
  \bmath{v}\cdot \bmath{\nabla v} = -\frac{\bmath{\nabla} P}{\rho}-\bmath{\nabla} 
\Phi \ ,
\label{euler}
\end{equation}
where $P$ is pressure; $\rho$ is density; $\Phi$ is the gravitational potential, 
which satisfies the Poisson equation
\begin{equation}
  \nabla^2 \Phi=4\pi G \rho \ ,
\label{poisson}
\end{equation}
where $G$ is the gravitational constant. The fluid's velocity $\bmath{v}$ is 
related to the rotational angular frequency $\Omega$ by 
$\bmath{v}=\Omega \varpi \bmath{e}_{\hat{\varphi}}$, where $\varpi$ is the distance 
from the rotation axis and $\bmath{e}_{\hat{\varphi}}$ is the unit vector along 
the azimuthal direction. The EOS we use is barotropic, i.e.\ 
$P=P(\rho)$, so the Euler equation 
can be integrated to give
\begin{equation}
  h=C-\Phi+{\varpi^2\over 2} \Omega^2 \ ,
\label{inteluer}
\end{equation}
where $C$ is a constant. The enthalpy (per mass) $h$ is given by
\begin{equation}
  h=\int_0^P {dP\over \rho} \ ,
\label{enthalpy}
\end{equation}
and is defined only inside the star. The boundary of the star is the surface 
with $h=0$.

The equilibrium configuration is determined by Hachisu's self-consistent field 
method \cite{hachisu86}: given an enthalpy distribution
$h_i$, we calculate the density distribution $\rho_i$ from the inverse of 
equation (\ref{enthalpy}) and from the EOS.
Next we calculate the gravitational potential $\Phi_i$ 
everywhere by solving the Poisson equation~(\ref{poisson}). Then the enthalpy 
is updated by 
\begin{equation}
  h_{i+1}=C_{i+1}-\Phi_{i}+{\varpi^2\over 2} \Omega_{i+1}^2 \ ,
\end{equation}
with $C_{i+1}$ and $\Omega_{i+1}^2$ determined by two boundary conditions. In Hachisu's 
paper \shortcite{hachisu86}, the axis ratio, i.e.\ the ratio of polar to equatorial 
radii, and the maximum density are fixed to 
determine $C_{i+1}$ and $\Omega_{i+1}^2$. However, we find it more convenient in 
our case to fix the equatorial radius $R_e$ and central enthalpy $h_c$, so that 
\begin{eqnarray}
  C_{i+1} &=& h_c+\Phi_i(0) \\
  \Omega_{i+1}^2 &=& -{2\over R_e^2} [C_{i+1}-\Phi_i(A)] \ ,
\end{eqnarray}
where $\Phi_i(0)$ and $\Phi_i(A)$ are the gravitational potential at the centre and 
at the star's equatorial surface respectively. The procedure is repeated
until the enthalpy and density distribution converge to the desired degree of 
accuracy. 

We used a spherical grid with $L$ radial spokes and $N$ evenly 
spaced grid points along each radial spoke. The spokes are located at angles 
$\theta_i$ in such a way that $\cos \theta_i$ correspond to the zeros of 
the Legendre polynomial of order $2L-1$: $P_{2L-1}(\cos \theta_i)=0$. 
Because of the reflection symmetry, we only need to consider spokes lying 
in the first quadrant. 
Poisson's equation is solved using the technique
described by Ipser and Lindblom~\shortcite{ipser90}.
The special choice of the angular positions of the radial spokes and the 
finite difference scheme make our numerical solution equivalent to an expansion 
in Legendre polynomials through order $l=2L-2$~\cite{ipser90}.
Although the white dwarfs we consider here are 
rapidly rotating, the equilibrium configurations are close to spherical, 
as demonstrated in the next subsection. So a relatively small number of 
radial spokes are adequate to describe the stellar models accurately.
We compared the results of $(L,N)=(10,3000)$ with 
$(L,N)=(20,5000)$ and find agreement to an accuracy of $10^{-5}$.
The accuracy of the model can also be measured by the virial theorem, which states 
that $2T_{\rm rot}+W+3\Pi=0$ for any equilibrium star (see e.g.\ Tassoul 1978). 
Here $T_{\rm rot}$ is the
rotational kinetic energy; $W$ is the gravitational potential energy and 
$\Pi =\int P\, d^3x$. We define 
\begin{equation}
  \epsilon= \left|\frac{2T_{\rm rot}+W+3\Pi}{W}\right| \ .
\label{VC:def}
\end{equation}
All models constructed in this section have $\epsilon \approx 10^{-7}$.

\subsection{Results}
\label{rtwd:result}

We constructed three models of rigidly rotating white dwarfs. All of them 
are maximally rotating: material at the star's equatorial surface rotates
at the local orbital frequency. Models~I and II are C-O white dwarfs with central 
densities $\rho_c=10^{10}~\rmn{g}~\rmn{cm}^{-3}$ and $\rho_c=6\times 
10^9~\rmn{g}~\rmn{cm}^{-3}$ respectively; Model~III is an O-Ne-Mg white 
dwarf with 
$\rho_c=4\times 10^9~\rmn{g}~\rmn{cm}^{-3}$. The properties of these 
white dwarfs are summarized in Table~\ref{wdmodels}. We see that the angular 
momentum $J$ decreases as the central density $\rho_c$ increases, because 
the white dwarf becomes smaller and more centrally condensed. Although
$T_{\rm rot}$ increases with $\rho_c$, $|W|$ increases at a faster rate so 
that $\beta=T_{\rm rot}/|W|$ decreases with increasing $\rho_c$. 
We also notice that the mass does not change 
much with increasing $\rho_c$. The reason is that massive white dwarfs are centrally 
condensed so their masses are determined primarily by the high density central 
core. Here the degenerate electron gas becomes highly relativistic
and the Coulomb effects are negligible, so the composition difference is 
irrelevant. Hence the white dwarf behaves 
like an $n=3$ polytrope, whose mass in the non-rotating case is independent 
of the central density. 

The masses of our three models are all greater than the Chandrasekhar 
limit for non-rotating white dwarfs. A non-rotating C-O white 
dwarf with $\rho_c=10^{10}~\rmn{g}~\rmn{cm}^{-3}$ has a radius 
$R=1300~\rmn{km}$ and a mass $M=1.40 M_{\odot}$. When this white dwarf is spun up 
to maximum rotation while keeping its mass fixed, the star puffs up to an oblate 
figure of equatorial radius $R_e=4100~\rmn{km}$ and polar radius $R_p=2700~\rmn{km}$, 
and its central density drops to $\rho_c=5.5\times 10^8~\rmn{g}~\rmn{cm}^{-3}$. 
This peculiar behaviour is caused by the soft EOS of relativistic 
degenerate electrons, 
which makes the star highly compressible and also highly expansible. 
When the angular velocity of the star is increased, the centrifugal 
force causes a large reduction in central density,  resulting in a dramatic increase 
in the overall size of the star.

\begin{table*}
\caption{The central density $\rho_c$, mass $M$, angular momentum $J$, 
rotational frequency $\Omega$, rotational kinetic energy $T_{\rm rot}$, the ratio of 
rotational kinetic to gravitational energies $\beta$, equatorial radius $R_e$ 
and polar radius $R_p$ of three rigidly and maximally rotating white dwarfs.}
\label{wdmodels}
\begin{tabular}{cccccccccc}
\hline
  & Composition & $\rho_c$ & $M$ & $J$ & $\Omega$ & $T_{\rm rot}$ & $\beta$ & $R_e$ & 
$R_p$ \\
& & $\rmn{g}~cm^{-3}$ &  $M_{\odot}$ & 
$\rmn{g}~\rmn{cm}^2~\rmn{s}^{-1}$ & $\rmn{rad}~\rmn{s}^{-1}$ & 
erg & & km & km \\
\hline
  Model I & C, O & $10^{10}$ & $1.47$ & $3.12\times 10^{49}$ & $5.37$ & 
$8.38\times 10^{49}$ & $0.015$ & $1895$ & $1247$ \\
  Model II & C, O & $6\times 10^9$ & $1.46$ & $3.51\times 10^{49}$ & $4.32$ & 
$7.57\times 10^{49}$ & $0.017$ & $2189$ &
$1439$ \\
  Model III & O, Ne, Mg & $4\times 10^9$ & $1.45$ & $3.80\times 10^{49}$ & $3.65$ & 
$6.94\times 10^{49}$ & $0.018$ & $2441$ & $1602$ \\
\hline
\end{tabular}
\end{table*}

\begin{figure}
\epsfig{file=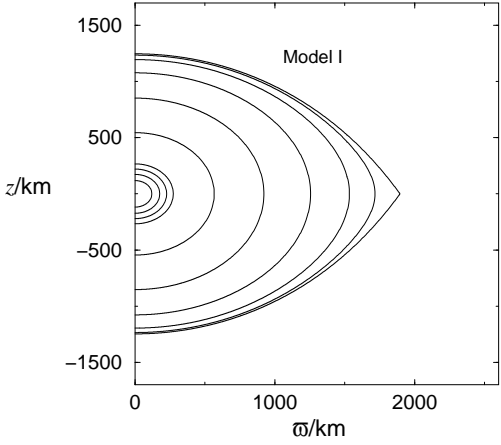,width=6cm,angle=270}
\caption{Meridional density contours of the rotating white dwarf 
of Model~I. The contours, from inward to outward, correspond to densities 
$\rho/\rho_c=$0.8, 0.6, 0.4, 0.2, 
0.1, $10^{-2}$, $10^{-3}$, $10^{-4}$, $10^{-5}$ and zero.}
\label{denWDI}
\end{figure}

\begin{figure}
\epsfig{file=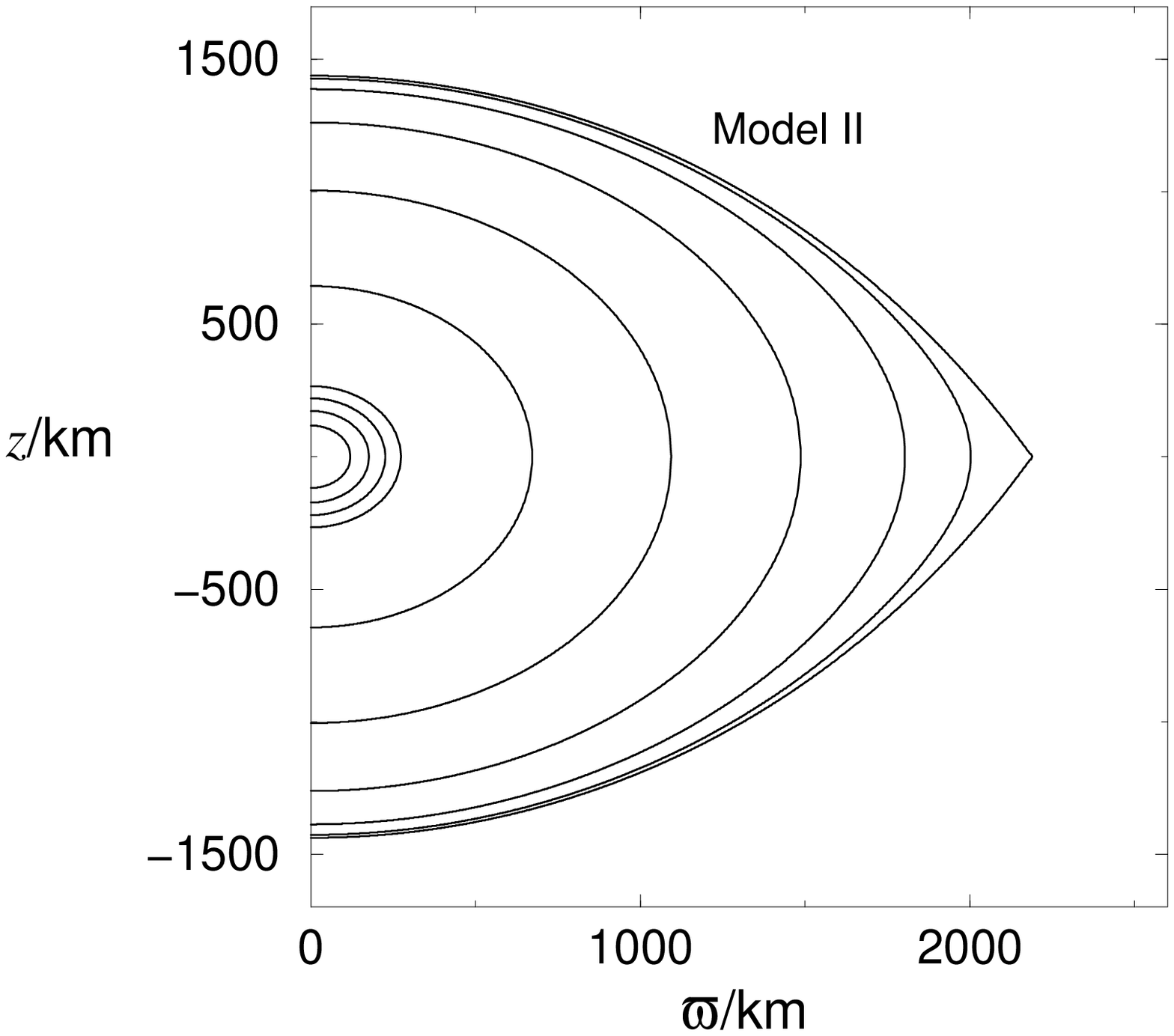,width=6cm,angle=270}
\caption{Same as Figure~\ref{denWDI} but for Model~II}
\label{denWDII}
\end{figure}

\begin{figure}
\epsfig{file=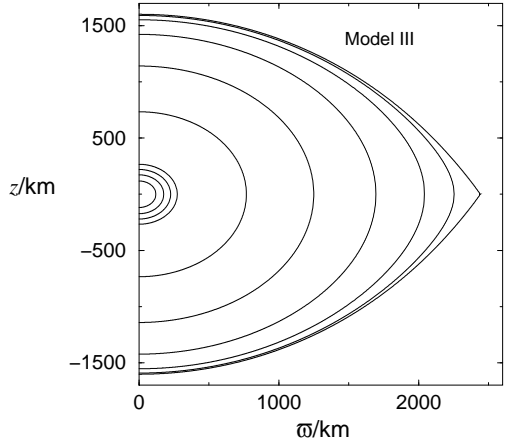,width=6cm,angle=270}
\caption{Same as Figure~\ref{denWDI} but for Model~III}
\label{denWDIII}
\end{figure}

\begin{figure}
\epsfig{file=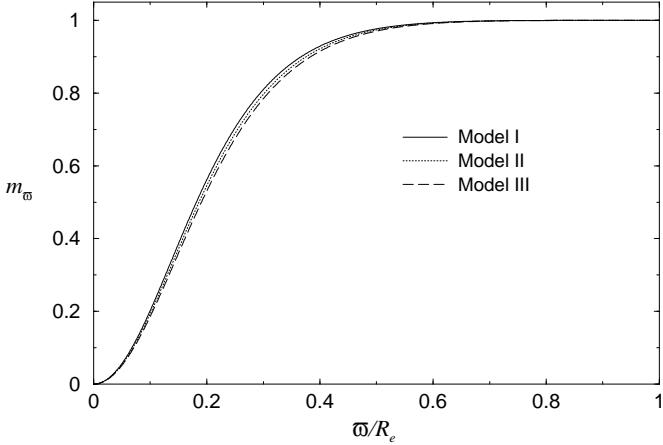,width=6cm,angle=270}
\caption{Cylindrical mass fraction $m_{\varpi}$ as a function of $\varpi$. 
Solid line corresponds to Model~I; dotted line, Model~II, and 
dashed line, Model~III.}
\label{fig:mpmWD}
\end{figure}

Figures~\ref{denWDI}-\ref{denWDIII} display the density contours of our
three models. The contours in the high density region remain more or less 
spherical even though our models represent the most rapidly rotating cases. 
The effect of rotation is only to flatten the density contours of the outer 
region in which the density is relatively low. This suggests that the white 
dwarfs are centrally condensed, and is clearly demonstrated in 
Figure~\ref{fig:mpmWD}, where the cylindrical mass fraction 
\begin{equation}
  m_{\varpi}=\frac{2\pi}{M}\int_0^{\varpi} d\varpi'\, \varpi'\, 
\int_{-\infty}^{\infty} dz'\, \rho(\varpi',z')
\label{m:def}
\end{equation}
is plotted. In all of our three models, more than half of the mass is concentrated 
inside the cylinder with $\varpi \approx 0.2 R_e$.

\begin{figure}
\epsfig{file=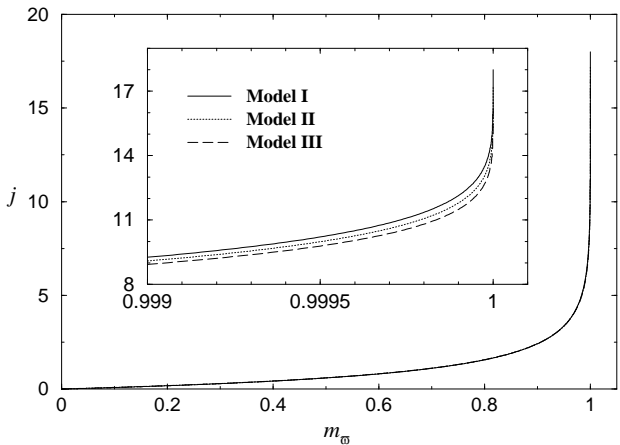,width=6cm,angle=270}
\caption{Normalized specific angular momentum $j$ as a function of the cylindrical 
mass fraction $m_{\varpi}$. The curves for the three models are indistinguishable 
except in the region very close to $m_{\varpi}=1$, which is magnified in the inset.}
\label{fig:jm}
\end{figure}

Figure~\ref{fig:jm} shows the specific angular momentum $j$ as a function of 
the cylindrical mass fraction $m_{\varpi}$, normalized so that $\int_0^1 
j(m_{\varpi})\, dm_{\varpi}=1$. 
The $j(m_{\varpi})$-curves for the three models are almost 
indistinguishable except in the region where $m_{\varpi}\approx 1$. The spike of the 
curve near $m_{\varpi}=1$ can be understood from Figure~\ref{fig:mpmWD}, where 
we see that $m_{\varpi}\approx 1$ when $\varpi/R_e \ga 0.6$. However, $j=(M/J)\Omega 
\varpi^2\, \propto \varpi^2$. These two make the values of $j$ in the interval 
$0.6^2 \la j/j(m_{\varpi}=1) \le 1$ squeeze to the region $m\approx 1$, and the 
spike results. 
We shall point out in the next section that this spike causes a serious numerical 
problem in the construction of the equilibrium models of the collapsed 
objects. The problem can be solved by truncating the upper part of 
the $j(m_{\varpi})$ curve. 
The physical justification is that the material in the outer 
region contributes
only a very small fraction of the total mass and angular momentum of the star, as 
illustrated in Table~\ref{outlayers} for Model~I. The situations for the other 
two models are very similar and so are not shown. We see that material in 
the region where $\varpi/R_e > 0.9$ [$j(m_{\varpi})/j(1)> 0.81$]
contributes less than $10^{-5}$ 
of the total mass and $10^{-4}$ of the total angular momentum. So the upper 
19 per cent of the $j(m_{\varpi})$-curve has little influence to the inner structure 
of the collapsed star. While this region is important for the structure of the 
star's outer layers, that part of the star is not of our primary interest 
since the mass there is too small to develop any instability that can 
produce a substantial amount of gravitational radiation.

\begin{table}
\caption{The outer layers of Model~I white dwarf. 
$J_{\varpi}$ is the angular momentum of the material inside the 
cylinder of radius $\varpi$.}
\label{outlayers}
\begin{tabular}{cccc}
\hline
 $\varpi/R_e$ & $1-m_{\varpi}$ & $1-J_{\varpi}/J$ & $j(m_{\varpi})/j(1)$ \\
\hline
 0.83 & $7.5\times 10^{-5}$ & $100\times 10^{-5}$ & 0.69 \\
 0.86 & $3.5\times 10^{-5}$ & $49\times 10^{-5}$ & 0.73 \\
 0.90 & $0.65\times 10^{-5}$ & $9.8\times 10^{-5}$ & 0.81  \\
 0.95 & $0.027\times 10^{-5}$ & $0.45\times 10^{-5}$ & 0.90 \\
\hline
\end{tabular}
\end{table}

\section{COLLAPSED OBJECTS}
\label{col-mod}

In this section, we present the equilibrium new-born neutron-star models 
that may result from the collapse of the three white dwarfs computed in the 
previous section. 
Instead of performing hydrodynamic simulations, we adopt a simpler approach:

First, we assume the collapsed stars are axisymmetric and are in rotational equilibrium 
with no meridional circulation. Second, we assume the EOS is barotropic, 
$P=P(\rho)$. These two assumptions imply that (1) the angular velocity $\Omega$ 
is a function of $\varpi$ only, i.e.\ $\partial \Omega/\partial z=0$, and 
(2) the Solberg condition is satisfied, which states that $dj/d\varpi>0$ for 
stable barotropic stars in rotational equilibrium (see 
e.g.\ Tassoul 1978). The angular velocity profile ($\partial \Omega/\partial z=0$) 
is observed in the 
simulations of M\"{o}nchmeyer, Janka and M\"{u}ller (M\"{o}nchmeyer \& 
M\"{u}ller 1988; Janka \& M\"{o}nchmeyer 1989ab). Third, we are only 
interested in the structure of the neutron stars within a few minutes after 
core bounce. The timescale is much shorter than any of the viscuous 
timescales, so viscosity does not have time to change the angular momentum of 
a fluid particle \cite{cutler87,sawyer89,lindblom79,horn81,goodwin82}. 
Finally, we assume no material is ejected during the collapse. 
It follows, from the conservation of $j$ and the fact that $j$ is 
a function of $\varpi$ only before and after collapse, that 
all particles initially located on a cylindrical 
surface of radius $\varpi_1$ from the rotation axis will end up being on a new 
cylindrical surface of radius $\varpi_2$. And the Solberg condition ensures that all 
particles initially inside 
the cylinder of radius $\varpi_1$ will collapse to the region inside the new 
cylinder of radius $\varpi_2$. Hence the specific angular momentum distribution 
$j(m_{\varpi})$ of the new equilibrium configuration is the same as that of the 
pre-collapse white dwarf; 
here $m_{\varpi}$ is the cylindrical mass fraction defined by equation~(\ref{m:def}).

Based on these assumptions, we constructed equilibrium models of the collapsed 
objects with the same masses, total angular momenta and $j(m_{\varpi})$ as the 
pre-collapse white dwarfs.

\subsection{Equations of state}

The gravitational collapse of a massive white dwarf is halted when the 
central density reaches nuclear density where the EOS becomes stiff. 
The core bounces back and within a few milliseconds, a hot ($T \ga 20~\rmn{Mev}$), 
lepton rich protoneutron star settles into hydrodynamic equilibrium.
During the 
so-called Kelvin-Helmholtz cooling phase, the temperature and lepton number 
decrease due to neutrino emission and the protoneutron star cools to a 
cold neutron star with temperature $T<1~\rmn{Mev}$ after several 
minutes. Since the cooling timescale is much longer than the hydrodynamical 
timescale, the protoneutron star can be regarded as in quasi-equilibrium.

The EOS of a protoneutron star is expressed in the form $P=P(\rho;s,Y_e)$, 
where $s$ and $Y_e$ are the entropy per baryon and lepton fraction respectively. 
As pointed out by Strobel, Scraab \& Weigel \shortcite{strobel99}, the structure 
of a protoneutron star can be approximated by a constant $s$ and $Y_e$ 
throughout the star, resulting in an effectively barotropic EOS. 

We used two different EOS for densities above 
$10^{10}~\rmn{g}~\rmn{cm}^{-3}$. The first is one of the standard EOS
for cold neutron stars. We adopt the Bethe-Johnson EOS 
\cite{bethe74} for densities above $10^{14}~\rmn{g}~\rmn{cm}^{-3}$, and BBP 
EOS \cite{baym71} for densities in the region 
$10^{11}~\rmn{g}~\rmn{cm}^{-3}\ - \ 10^{14}~\rmn{g}~\rmn{cm}^{-3}$.
It turns out that the densities of these collapsed stars are lower than 
$4\times 10^{14}~\rmn{g}~\rmn{cm}^{-3}$, and ideas about the EOS in this 
range have not changed very much since 1970's.
The second is the EOS $\rmn{LPNS}_{\rm YL04}^{\rm s2}$ of Strobel et al 
\shortcite{strobel99}\footnote{The tabulated EOS is obtained from
http://www.physik.uni-muenchen.de/sektion/suessmann/astro/eos/.}. 
This corresponds to a 
protoneutron star $0.5 - 1~\rmn{s}$ after core bounce. It has an entropy per 
baryon $s=2k_{\rm B}$ and a lepton fraction $Y_e=0.4$, where $k_{\rm B}$ 
is Boltzmann's constant. 
We join both EOS to that of the pre-collapse white dwarf for
densities below $10^{10}~\rmn{g}~\rmn{cm}^{-3}$. Hereafter, we shall 
call the first EOS the cold EOS, and the second one, the hot EOS.

\subsection{Numerical method}
\label{col:numerical}

We compute the equilibrium structure by Hachisu's self-consistent field 
method modified so that $j(m_{\varpi})$ can be specified \cite{smith92}. The 
iteration scheme is based on the integrated static Euler equation~(\ref{euler}) 
written in the form 
\begin{equation}
  h(\varpi,z) = C-\Phi(\varpi,z) + \left({J\over M}\right)^2
\int_0^{\varpi} d\varpi' \frac{j^2(m_{\varpi'})}{\varpi'^3} \ ,
\label{int:eluer2}
\end{equation}
where $C$ is the integration constant, and $M$ and $J$ are the total mass and 
angular momentum of the star respectively.
Given an enthalpy distribution 
$h_i$ everywhere, the density distribution $\rho_i$ is calculated by the 
EOS and the inverse of equation~(\ref{enthalpy}).
Next we compute the mass $M_i$ and cylindrical mass fraction $m_{\varpi,i}$ by
\begin{eqnarray}
  M_i &=& 4\pi \int_0^{\infty}d\varpi' \varpi' \int_0^{\infty}dz'
\rho_i(\varpi',z') \cr
  m_{\varpi,i} &=& \frac{4\pi}{M_i}\int_0^{\varpi}d\varpi' \varpi' 
\int_0^{\infty}dz' \rho_i(\varpi',z') \ ,
\label{m:cal}
\end{eqnarray}
and solve the Poisson equation $\nabla^2 \Phi_i=4\pi G \rho_i$ to obtain the 
gravitational potential $\Phi_i$. We then update the enthalpy by 
equation~(\ref{int:eluer2}):
\begin{eqnarray}
  h_{i+1}(\varpi,z) &=& C_{i+1}-\Phi_i(\varpi,z) \cr
 & & + \left({J_{i+1}\over M_{i+1}}\right)^2
\int_0^{\varpi} d\varpi' \frac{j^2(m_{\varpi',i})}{\varpi'^3} \ ,
\label{h:cal}
\end{eqnarray}
with the parameters $C_{i+1}$ and $(J_{i+1}/M_{i+1})^2$ determined by specifying 
the central density $\rho_c$ and equatorial radius $R_e$. The procedure is 
repeated until the enthalpy and density distribution converge to the desired 
degree of accuracy.

To construct the equilibrium configuration with the same total mass and angular 
momentum as a pre-collapse white dwarf, we first compute a model of a non-rotating 
spherical neutron star, use its enthalpy distribution as an initial guess 
for the iteration scheme described above and build a configuration with 
slightly different $\rho_c$ or $R_e$. Then the parameters $\rho_c$ and 
$R_e$ are adjusted until we end up with a configuration having the correct total 
mass and angular momentum. 

Two numerical problems were encountered in this procedure. 
The first problem is that when the angular momentum $J$ is increased, the 
star becomes flattened, and the iteration often oscillates among two or more 
states without converging. This problem can be solved by using a revised 
iteration scheme suggested by Pickett, Durisen \& Davis \shortcite{pickett96}, 
in which only a fraction of the revised enthalpy $h_{i+1}$, 
$h'_{i+1}=(1-\delta)h_{i+1}+\delta h_{i}$, is used for 
the next iteration. Here $\delta<1$ is a parameter controlling the change 
of enthalpy. We need to use $\delta > 0.95$ for very flattened configurations, 
and it takes $100 - 200$ iterations for the enthalpy and density distributions 
to converge.

The second problem has to do with the spike of the $j(m_{\varpi})$-curve near 
$m_{\varpi}=1$ (see Figure~\ref{fig:jm}). The slope is so steep that it makes 
the iteration unstable. As discussed in Section~\ref{rtwd:result}, the material 
in the region very close to $m_{\varpi}=1$ contains a very small amount of mass 
and angular momentum, so we can truncate the last part of 
the $j(m_{\varpi})$-curve without introducing much error. Specifically, we set a
parameter $j_c<j(m_{\varpi}=1)$, compute a quantity $m_c$ which satisfies 
$j(m_c)=j_c$. Then we use the specific angular momentum distribution 
$\tilde{j}(m_{\varpi})=j(m_{\varpi}\, m_c)$ instead of $j(m_{\varpi})$. Typically, 
we choose 
$j_c/j(1)=0.81$ so that $1-m_c \approx 10^{-5}$ (see Table~\ref{outlayers}). 
Hence the distributions $\tilde{j}(m_{\varpi})$ and $j(m_{\varpi})$ are basically 
the same except in the star's outermost region, which is unimportant to the inner
structure the star, and presumably also unimportant for the star's dynamical 
and secular stabilities. We also tried several different values of $j_c$ and found that 
the change of physical properties of the collapsed objects (e.g.\ the quantities 
in Table~\ref{tab:prop}) are within the error due to 
our finite-size grid. Thus the truncation is also justified numerically.

We evaluate these stellar models on a cylindrical grid. This allows us to 
compute the integrals in equations~(\ref{m:cal}) and (\ref{h:cal}) easily. We find it 
more convenient however, to solve the the Poisson equation for the gravitational 
potential on a spherical grid using the method described by Ipser and Lindblom 
\shortcite{ipser90}. We have verified that the potential obtained in this 
way agrees with the result obtained with a cylindrical multi-grid solver to 
within $0.5\%$. However, the spherical grid solver (including the needed 
transformation from one grid to the other) is much faster than the cylindrical 
grid solver. The accuracy of our final equilibrium models can also be measured 
by the quantity $\epsilon$ defined in equation~(\ref{VC:def}).
The values of $\epsilon$ for models computed in this section are few times $10^{-4}$.

\subsection{Results}

\begin{table}
\caption{The central density $\rho_c$, radius of gyration $R_g$, characteristic 
radius $R_*$ and ratio of rotational kinetic energy to gravitational energy 
$\beta$ of the collapsed objects with the cold and the hot 
EOS.}
\label{tab:prop}
\begin{tabular}{ccccc}
\hline
 & $\rho_c$ & $R_g$ & $R_*$ & $\beta$ \\
 & $\rmn{g}~\rmn{cm}^{-3}$ & km & km & \\
\hline
 Model I (cold EOS) & $3.7\times 10^{14}$ & 63 & 670 & 0.230 \\
 Model I (hot EOS) & $1.4\times 10^{14}$ & 67 & 650 & 0.139 \\
 Model II (cold EOS) & $3.5\times 10^{14}$ & 78 & 800 & 0.246 \\
 Model II (hot EOS) & $0.79\times 10^{14}$ & 85 & 800 & 0.137 \\
 Model III (cold EOS) & $3.2\times 10^{14}$ & 94 & 940 & 0.261 \\
 Model III (hot EOS) & $0.27\times 10^{14}$ & 110 & 940 & 0.127 \\
\hline
\end{tabular}
\end{table}

Table~\ref{tab:prop} shows some properties of the collapsed 
objects resulting from the collapse of the three white dwarfs in 
Section~\ref{ini-mod}. We define the radius 
of gyration, $R_g$, and the characteristic radius, $R_*$, of the star by 
\begin{eqnarray}
 M R_g^2 &=& \int \rho \varpi^2 \, d^3 x \\
  m_{\varpi}(\varpi=R_*) &=& 0.999 \ .
\end{eqnarray}
We see that $R_g$ and $R_*$ that result from the same initial white dwarfs are 
insensitive to the neutron-star EOS, while 
there is a dramatic difference in the central density $\rho_c$ and 
the ratio of rotational kinetic energy to gravitational potential energy $\beta$. 
The collapsed stars with the hot EOS have smaller $\rho_c$ and 
$\beta$ than those with the cold EOS. 
In fact, the central densities of these hot stars are less than nuclear density.
It is well-known 
that a non-rotating star cannot have a central density in the sub-nuclear 
density regime ($4\times 10^{11}~\rmn{g}~\rmn{cm}^{-3}\la \rho \la
2\times 10^{14}~\rmn{g}~\rmn{cm}^{-3}$)
because the EOS is too soft to render the star stable against 
gravitational collapse. It has been suggested that if rotation is 
taken into account, a star with a central density in this regime can exist. 
Such stars are termed `fizzlers' in the literature 
\cite{shapiro76,tohline84,eriguchi85,muller85,hayashi98,imamura00}. However, these 
so-called fizzlers in our case can exist for only about 20~s before evolving
to rotating cold neutron stars. In order to build a stable cold model in the
sub-nuclear density regime, the collapsed star has to rotate much faster, which is 
impossible unless the pre-collapse white dwarf is highly differentially rotating.

We mention in Section~\ref{introduction} that Zwerger and M\"{u}ller \shortcite{zwerger97}
performed 2D hydrodynamic simulations of axisymmetric rotational core collapse.
Their pre-collapse models are rotating stars with $n=3$ polytropic EOS, which is 
close to the real EOS of a massive white dwarf.
All of their pre-collapse models have a
central density of $10^{10}~\rmn{g}~\rmn{cm}^{-3}$ (see their Table~1). The model A1B3
in their paper is the fastest (almost) rigidly rotating star, but its total angular
momentum $J$ and $\beta$ are respectively 22 per cent and 40 per cent less than 
those of our Model~I of the pre-collapse white dwarf, though both have the same 
central density. This suggests that the structure of a massive white dwarf 
is senstive to the EOS.
Zwerger and M\"{u}ller state in their paper that no equilibrium configuration
exists that has $\beta>0.01$ for the (almost) rigidly rotating case. This assertion
is confirmed by our numerical code. Zwerger and M\"{u}ller
adopt a simplified analytical EOS for the collapsing 
core. At the end of their simulations, the models A1B3G1-A1B3G5, corresponding to 
the collapsed models of A1B3, have values of $\beta$ 
less than 0.07, far smaller than the $\beta$'s of our collapsed model I 
(see Table~\ref{tab:prop}), indicating that the EOS of the collapsed 
objects also play an important role on the final equilibrium configurations (or that 
their analysis violates one of our assumptions).

\begin{figure*}
\epsfig{file=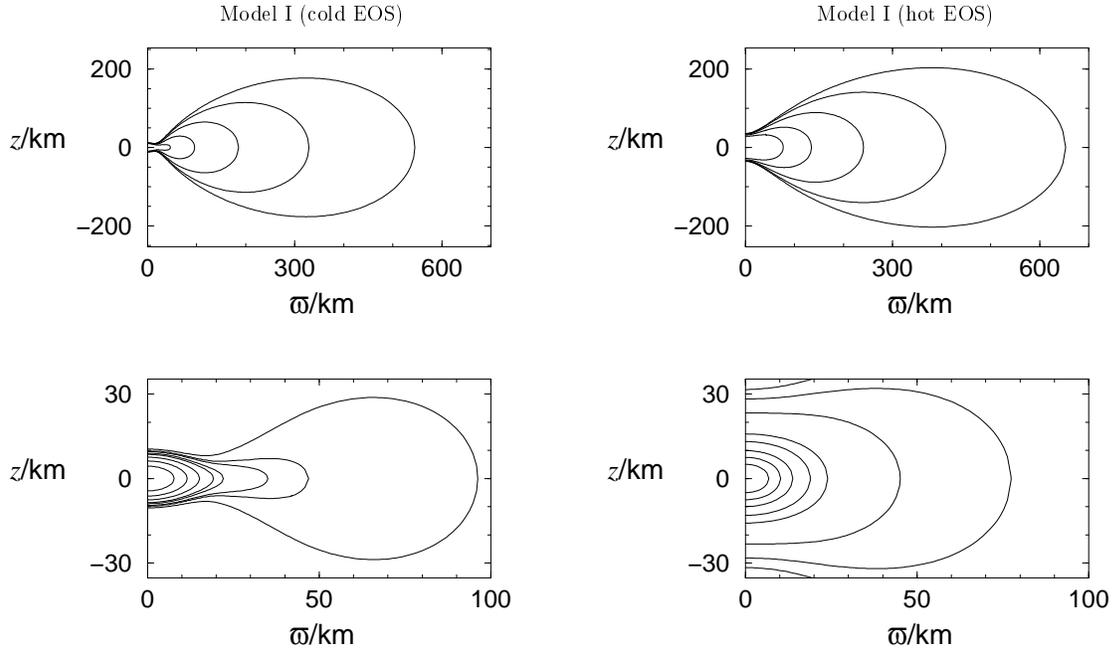,width=10cm,angle=270}
\caption{Meridional density contours of the neutron stars resulting from the
 collapse of Model~I white dwarf. 
The left graphs correspond to the cold EOS, and the right graphs, 
the hot EOS.
The contours in the upper graphs denote, from inward to outward, 
$\rho/\rho_c=10^{-3},\ 10^{-4},\ 10^{-5},\ 10^{-6}$, and $10^{-7}$. 
The contours in the lower graphs denote, from inward to outward,
$\rho/\rho_c=$0.8, 0.6, 0.4, 0.2, 0.1, $10^{-2}$, $10^{-3}$ and $10^{-4}$.}
\label{denI}
\end{figure*}

\begin{figure*}
\epsfig{file=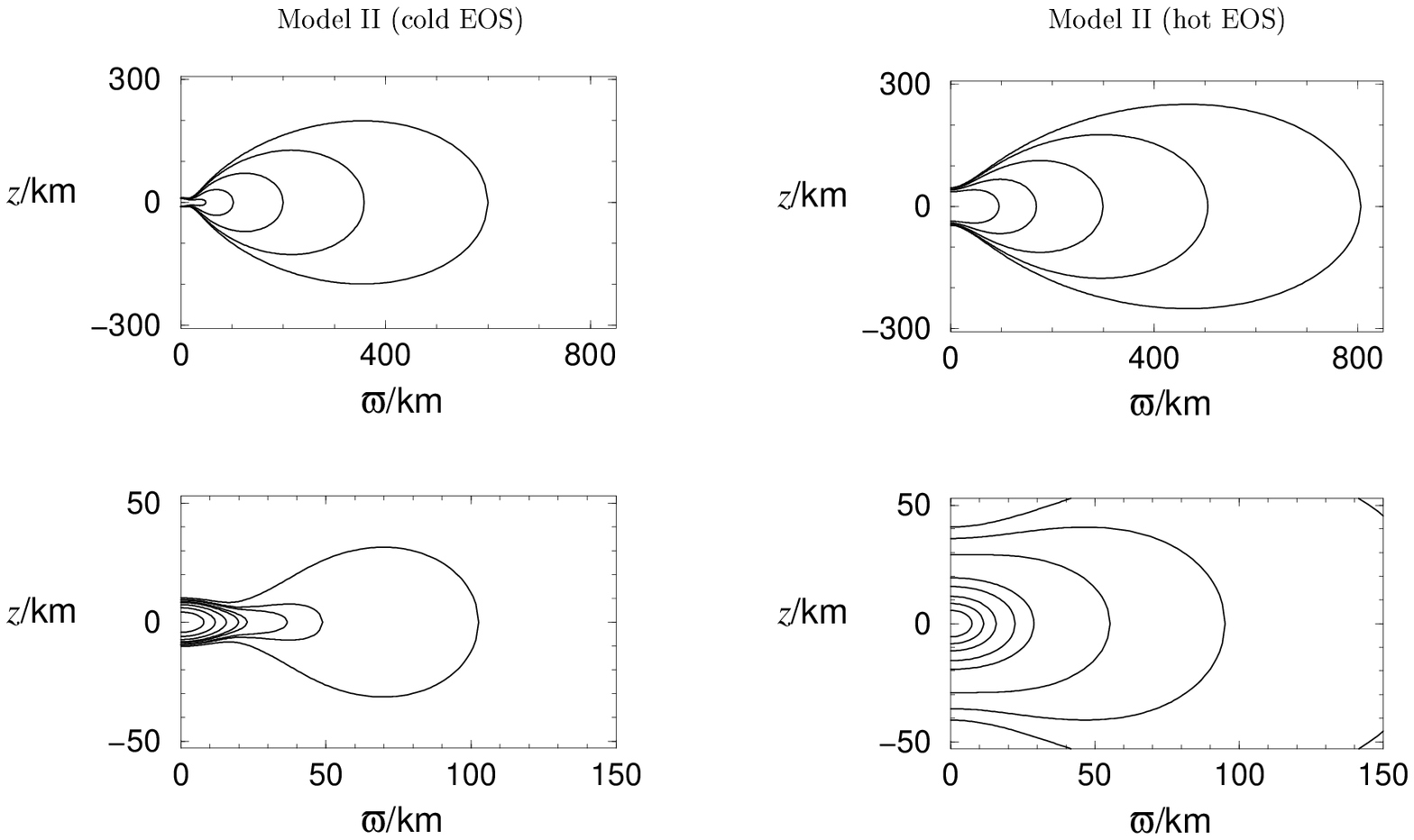,width=10cm,angle=270}
\caption{Same as Figure~\ref{denI} but for Model~II.}
\label{denII}
\end{figure*}

\begin{figure*}
\epsfig{file=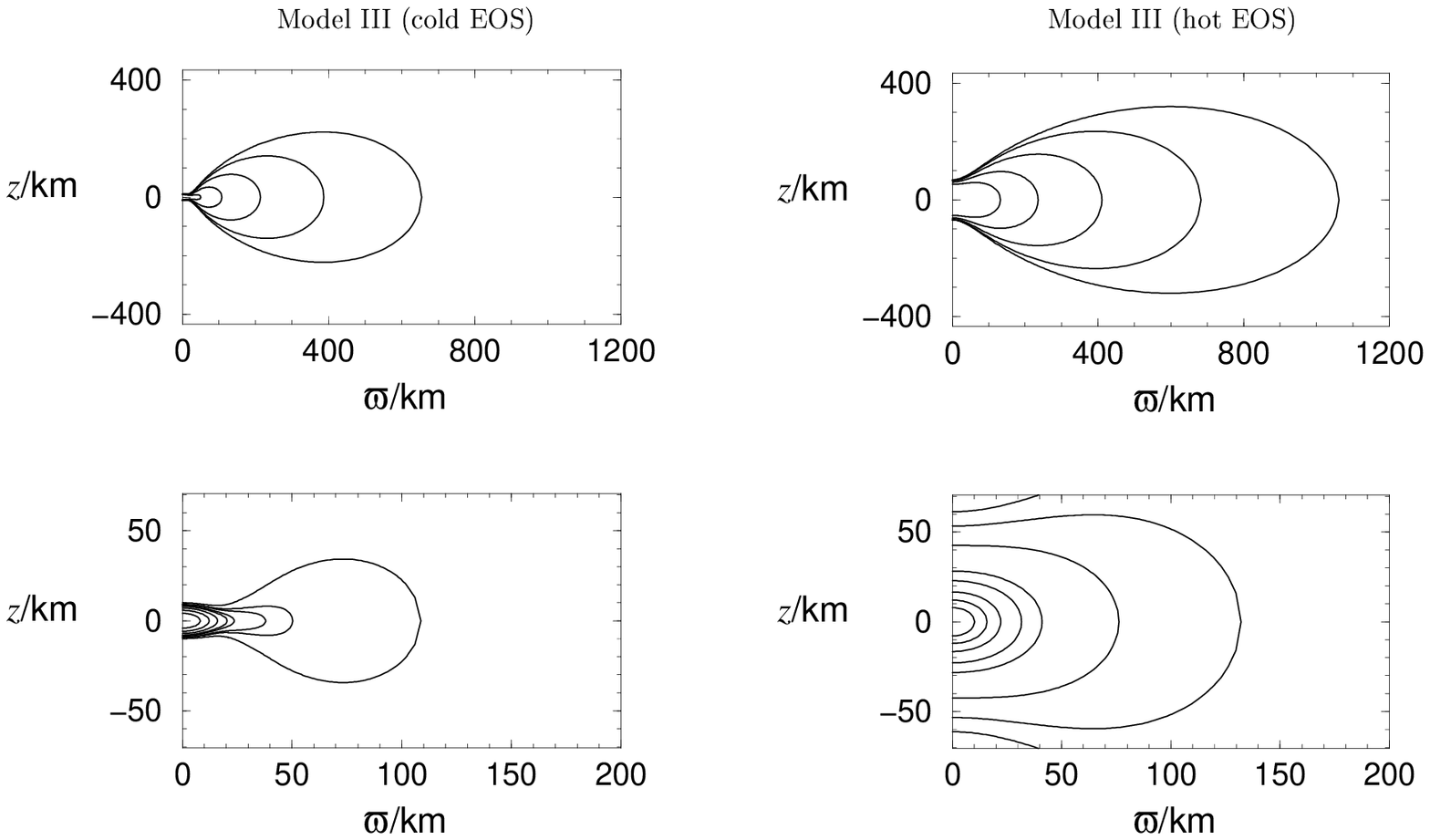,width=10cm,angle=270}
\caption{Same as Figure~\ref{denI} but for Model~III.}
\label{denIII}
\end{figure*}

Figures~\ref{denI}-\ref{denIII} show the density contours of the collapsed 
objects. We see that the contours of the dense central region look like the 
contours of a typical rotating star. As we go out to the low density region, the shapes 
of the contours become more and more disk-like. Eventually, the contours
turn into torus-like shapes for densities lower than $10^{-4}\rho_c$.
In all cases, the objects contain two regions: a dense 
central core of size about 20~km and a low density torus-like envelope 
extending out to 1000~km from the 
rotation axis. Since we truncate the $j(m_{\varpi})$-curve, we cannot determine 
accurately the actual boundary of the stars. The contours shown in these 
figures have been checked to move less than one per cent as the cutoff $j_c/j(1)$ 
is changed from 0.7 to 0.9. 
This little change is hardly visible at the displayed scales.

\begin{figure}
\begin{center}
\epsfig{file=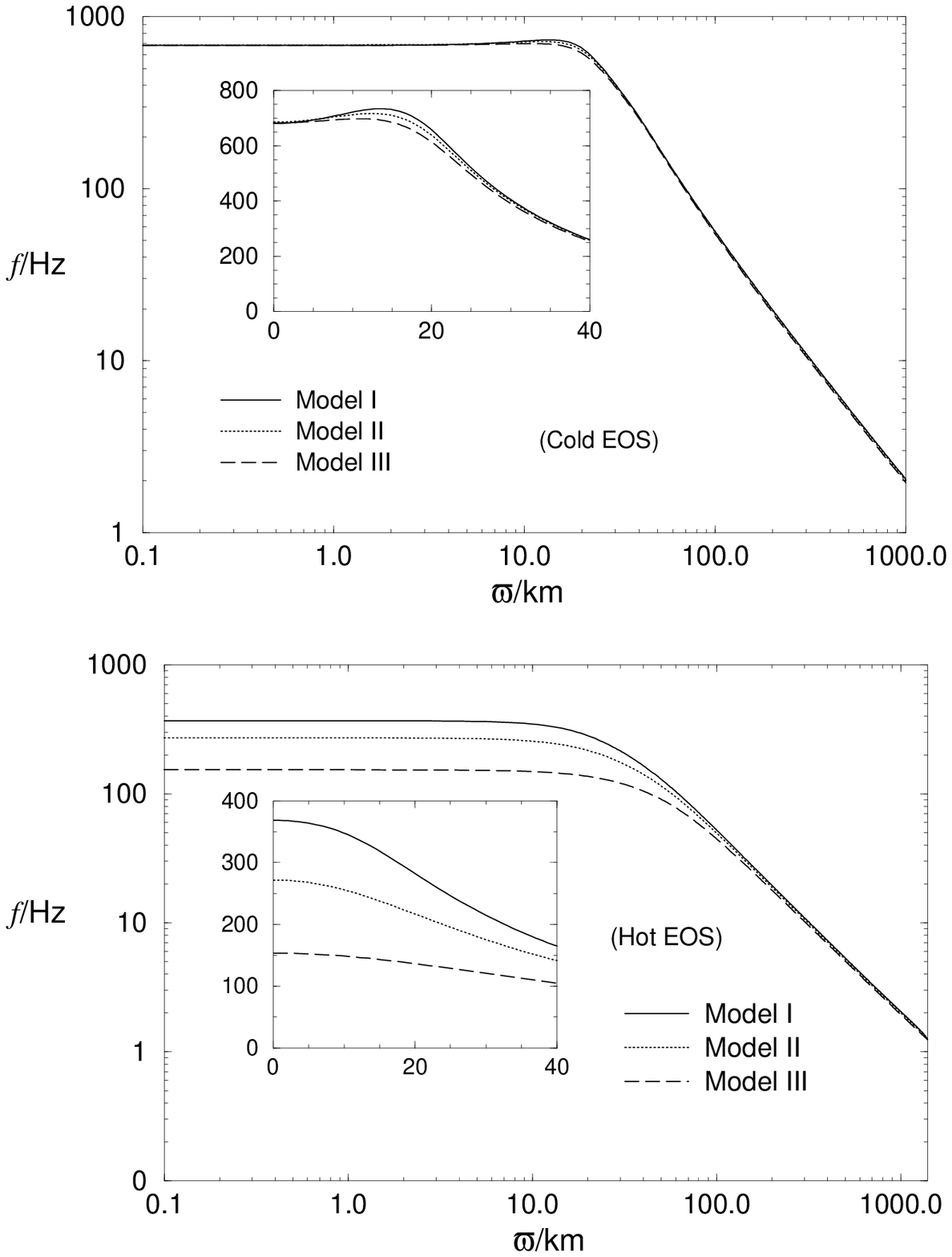,width=9cm}
\end{center}
\caption{Rotational frequency $f$ as a function $\varpi$ for the cold models 
(upper graph) and the hot models (lower graph). The inset 
in each graph shows $f$ in linear scale in the central region.}
\label{freq}
\end{figure}

Figure~\ref{freq} shows the rotational frequency $f\equiv \Omega/2\pi$ as
a function of $\varpi$, the distance from the rotation axis. We see
that the cores of the cold models are close to rigid rotation. The rotation
periods of the cores of the cold neutron stars are all about 1.4~ms, slightly 
less than the period of the fastest observed millisecond pulsar (1.56~ms).
A further analysis reveals that $f \propto \varpi^{-\alpha}$ in the region 
$\varpi \ga 100~\rmn{km}$, where $\alpha \approx 1.5$ for the cold models and 
$\alpha \approx 1.4$ for the hot models.

\begin{figure}
\begin{center}
\epsfig{file=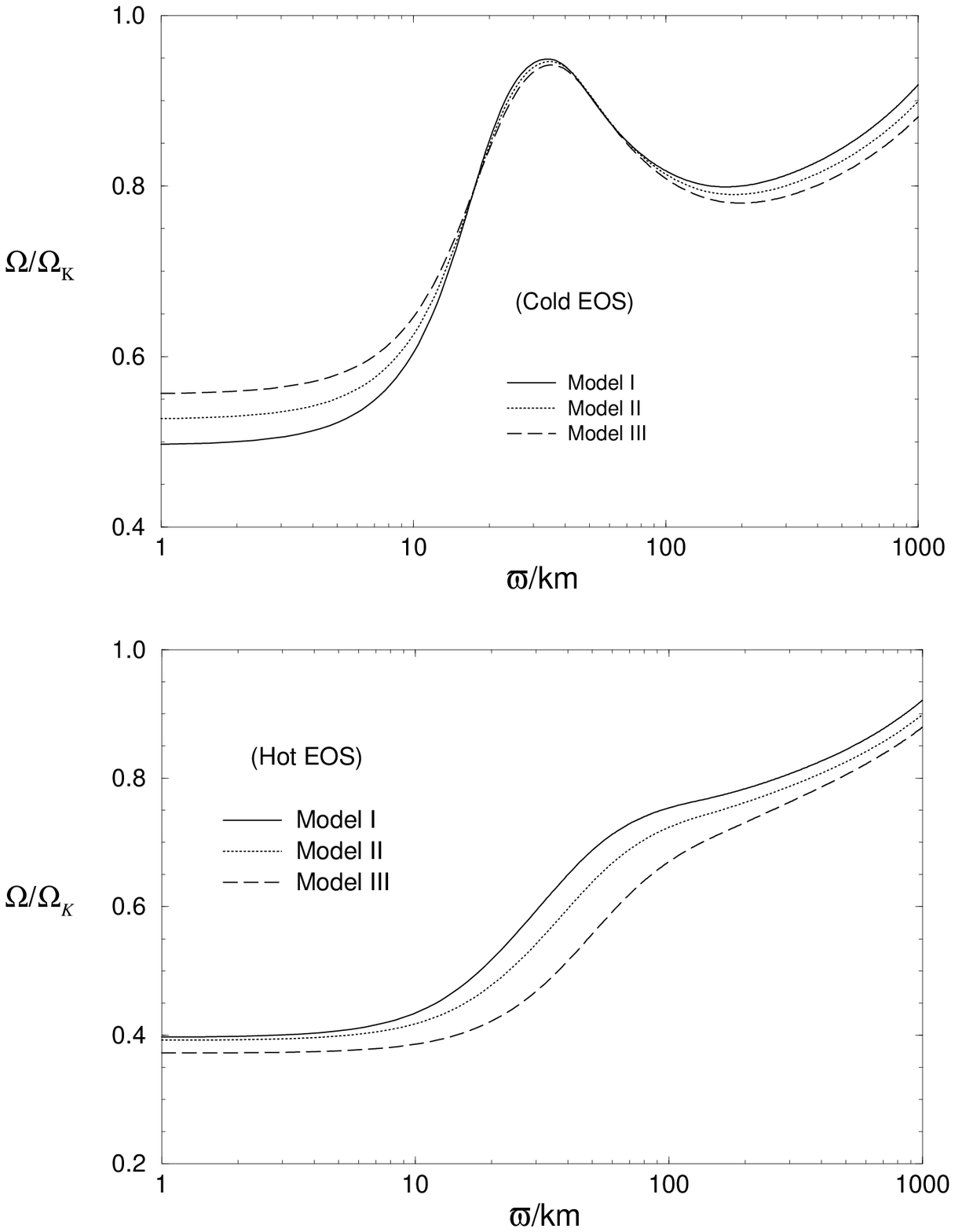,width=9cm}
\end{center}
\caption{The ratio $\Omega/\Omega_K$ along the equator as a function of $\varpi$ 
for the cold models (upper graph) and the hot models (lower graph).}
\label{freq-ratio}
\end{figure}

To gain an insight into the structure of the envelope, we define the Kepler 
frequency $\Omega_K$ at a given point on the equator as the angular frequency 
required for a particle to be completely supported by centrifugal force, 
i.e.\ $\Omega_K$ satisfies 
the equation $\Omega_K^2 \varpi=g$, where $g$ is the magnitude of gravitational 
acceleration at 
that point. Figure~\ref{freq-ratio} plots $\Omega/\Omega_K$ as a function of 
$\varpi$ along the equator. For the cold models, the curves increase 
from 0.5 at the centre to a maximum of about 0.95 at $\varpi \approx 35~\rmn{km}$, 
then decrease to a local minimum of about $0.8$, and then gradually 
increase in the outer region. The curves of the hot models, on the other hand, 
increase monotonically from about 0.4 at the centre to over 0.7 in the outer 
region. In all cases, centrifugal force plays an important role in the structure 
of the stars, especially in the low density region. 

\begin{figure}
\begin{center}
\epsfig{file=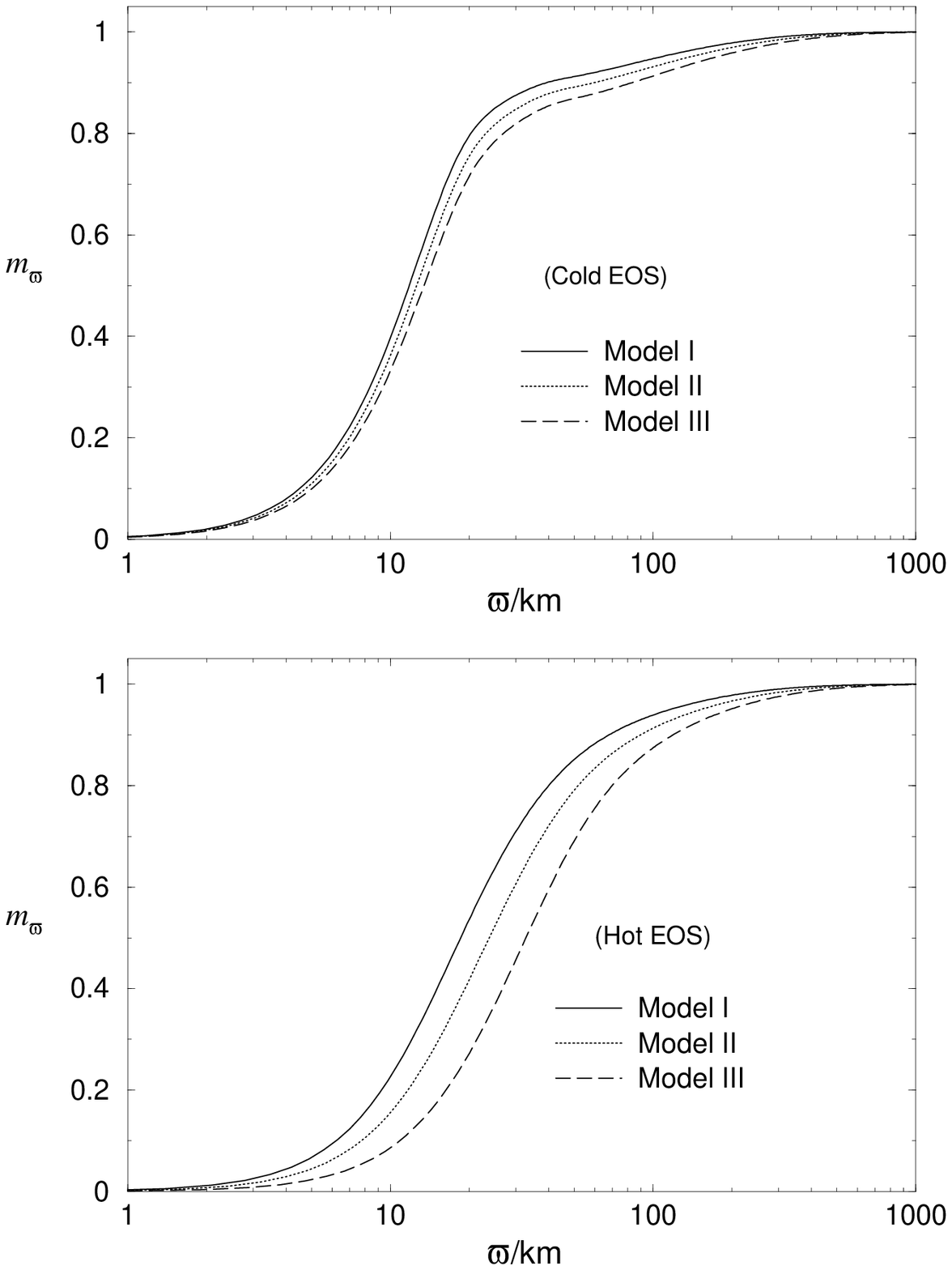,width=9cm}
\end{center}
\caption{Cylindrical mass fraction $m_{\varpi}$ as a function of $\varpi$ for 
the cold models (upper graph) and the hot models (lower graph).}
\label{fig:mpm}
\end{figure}

Figure~\ref{fig:mpm} plots the cylindrical mass fraction $m_{\varpi}$ as a 
function of $\varpi$. In all cases the cores contain most of the stars' mass. 
Material in 
the region $\varpi \ga 200~\rmn{km}$ occupies only a few per cent of the total
mass, but it is massive enough that its self-gravity cannot be neglected 
in order to compute the structure of the envelope accurately. 
The envelope can be regarded as a massive, self-gravitating accretion torus.
The same structure is also observed in the core collapse simulations of 
Janka \& M\"{o}nchmeyer \shortcite{janka89b} and Fryer \& Heger (private 
communication with Fryer).

\begin{figure}
\begin{center}
\epsfig{file=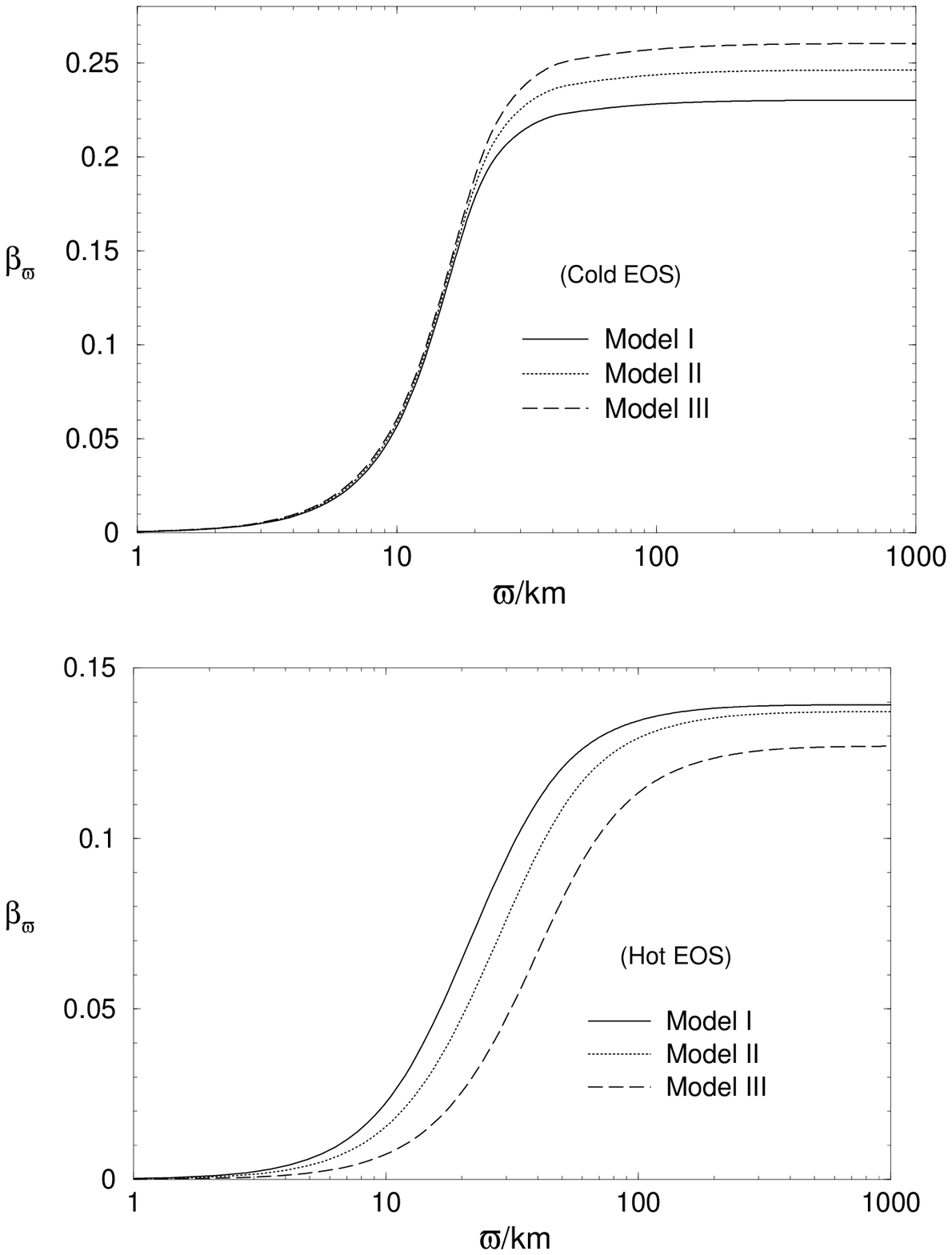,width=9cm}
\end{center}
\caption{The quantity $\beta_{\varpi}$ as a function of $\varpi$ for the 
cold models (upper graph) and the hot models (lower graph).}
\label{ToW}
\end{figure}
Figure~\ref{ToW} shows $\beta_{\varpi}=T_{\rm rot}(\varpi)/|W(\varpi)|$ as 
a function of $\varpi$, where $T_{\rm rot}(\varpi)$ and $W(\varpi)$ 
are the rotational 
kinetic energy and gravitational potential energy inside the cylinder of 
radius $\varpi$, i.e.
\begin{eqnarray}
  T_{\rm rot}(\varpi) &=& 2\pi \int_0^{\varpi} d\varpi' \varpi' (\Omega \varpi')^2 
\int_0^{\infty} dz' \rho(\varpi',z') \\
  W(\varpi) &=& 2\pi \int_0^{\varpi} d\varpi' \varpi' \int_0^{\infty} dz' 
\rho(\varpi',z') \Phi(\varpi',z') \ .
\end{eqnarray}
The values of $\beta_{\varpi}$ approach $\beta$ when $\varpi \ga 40~\rmn{km}$ for 
the cold EOS models and when $\varpi \ga 100~\rmn{km}$ for the hot EOS models. 
This suggests that material in the region $\varpi \ga 100~\rmn{km}$ contains 
negligible amount of kinetic energy, and any instability developed in 
this region could not produce strong gravitational waves.

\section{STABILITY OF THE COLLAPSED OBJECTS}
\label{stability}

We first consider axisymmetric instabilities, i.e. axisymmetric collapse. 
This stability is verified when we construct the models. Recall 
that we start from the model of a non-rotating spherical star which is
stable. Then we use it as an initial guess to build a sequence of rotating 
stellar models with the same specific angular momentum distribution but 
different total masses and angular momenta. If the final model we end up 
with is unstable against axisymmetric perturbations, there must be at least 
one model in the sequence such that 
\begin{equation}
  q \equiv \left. \frac{\partial M}{\partial \rho_c}\right|_{j(m_{\varpi}),J}=0 \ ,
\label{axis:cond}
\end{equation}
which signals the onset of instability \cite{bisnovatsyi74}.
Here $M$ is the total mass and $\rho_c$ is the central density. The partial 
derivative is evaluated by keeping the total angular momentum $J$ and specific 
angular momentum distribution $j(m_{\varpi})$ fixed. We have verified that 
all of our equilibrium models in the sequencies satisfy $q>0$. Hence they are 
all stable against axisymmetric perturbations.

We next consider non-axisymmetric instabilities. We have 
$\beta=0.23 - 0.26$ for the cold EOS models and $\beta=0.13 - 0.14$ for 
the hot EOS models (Table~\ref{tab:prop}). 
The hot models are probably dynamically 
stable but may be secularly unstable. However, since they are evolving to 
cold neutron stars in about 20~s and their structures are continually changing 
on times comparable to the secular timescale, we shall not discuss secular 
instabilities of these hot models here.

The values of $\beta$ for the three cold neutron stars are slightly less than the 
traditional critical value for dynamical instability, $\beta_d \approx 0.27$.
This critical value is based on simulations of differentially rotating 
polytropes having the $j(m_{\varpi})$-distribution of Maclaurin spheroids.
However, recent simulations demonstrate that differentially rotating polytropes 
having other $j(m_{\varpi})$-distributions can be dynamically unstable for
values of $\beta$ as low as 0.14 \cite{pickett96,centrella00}.
The equilibrium configurations of some of those unstable stars also contain
a low density accretion disk like structure in the stars' outer layers. This 
feature is very similar to the equilibrium structure of our models. 
Hence a more 
detailed study has to be carried out to determine whether the cold models are
dynamically stable. 

The subsequent evolution of a bar-unstable object has been studied for the
past 15 years 
\cite{durisen86,williams88,houser96,pickett96,smith96,houser98,new99,imamura00,brown00}. 
It is found that a bar-like structure develops in a dynamical timescale. 
However, it is still not certain whether the bar structure would be persistent, 
giving rise to 
a long-lived gravitational wave signal, or material would be shed from the 
ends of the bar after tens of rotation periods, leaving an axisymmetric, 
dynamically bar-stable central star.

Even if the cold neutron stars are dynamically stable, they are subject to 
various secular instabilities. The timescale of the gravitational-wave-driven
bar-mode instability can be estimated by 
\begin{equation}
  \tau_{\rm bar}=0.1~\rmn{s}\, \left(\frac{R}{35~\rmn{km}}\right)^{-5}
\left(\frac{\Omega}{4000~\rmn{rad}~\rmn{s}^{-1}}\right)^{-6} 
\left(\frac{\beta-\beta_s}{0.1}\right)^{-5} 
\end{equation}
\cite{friedman7578}. In our case, $R\approx 35~\rmn{km}$ (see Figure~\ref{ToW}), 
$\Omega\approx 4000~\rmn{rad}~\rmn{s}^{-1}$ and $\beta \approx 0.24$, 
so $\tau_{\rm bar} \sim 0.1~\rmn{s}$. Gravitational waves may also drive 
the r-mode instability \cite{lindblom98}. 
The timescale is estimated by 
\begin{equation}
  \tau_{\rm r} = 7.3~\rm{s}\ \left( \frac{\bar{\rho}}{10^{14}~\rmn{g}~\rmn{cm}^{-3}}
\right)^3 \left( \frac{\Omega}{4000~\rmn{rad}~\rmn{s}^{-1}}\right)^{-6}
\end{equation}
for the $l=2$ r-mode at low temperatures \cite{lindblom99}, where $\bar{\rho}$ 
is the average density. Inserting
$\bar{\rho}$ for the inner 20-km cores of the cold stars, 
we have $\tau_r \approx 10~\rmn{s}\gg \tau_{\rm bar}$. 
The evolution of the bar-mode secular instability has only been studied in 
detail for the Maclaurin spheroids. These objects evolve through a sequence 
of deformed non-axisymmetric configurations eventually to settle down as a 
more slowly rotating stable axisymmetric star \cite{lindblom77,lai95}. 
It is generally expected that stars having more realistic EOS
will behave similarly.

\section{CONCLUSIONS}
\label{conclusion}

We have constructed equilibrium models of differentially rotating neutron 
stars which model the end products of the accretion induced collapse 
of rapidly rotating white dwarfs. We considered three models for the 
pre-collapse white dwarfs. All of them are rigidly rotating at the 
maximum possible angular velocities. The white dwarfs are described by the 
EOS of degenerate 
electrons at zero temperature with Coulomb corrections derived by Salpeter 
\shortcite{salpeter61}.

We assumed that (1) the collapsed objects are axisymmetric and are in 
rotational equilibrium with no meridional circulation, (2) the EOS is 
barotropic, (3) viscosity can be neglected, and (4) any ejected material 
carries negligible amounts of mass and angular momentum. We then built 
the equilibrium models of the collapsed stars based on the fact that 
their final configurations must have the same masses,
total angular momenta and specific angular momentum distributions, $j(m_{\varpi})$,
as the pre-collapse white dwarfs.

Two EOS have been used for the collapsed objects. One of 
them is one of the standard cold neutron-star EOS. 
The other is a hot EOS
suitable for protoneutron stars, which are characterized by their 
high temperature and high lepton fraction. 

The equilibrium structure of the collapsed objects in all of our models consist 
of a high density central core of size about 20~km, surrounded by a massive 
accretion torus extending over 1000~km from the rotation axis.
More than 90 per cent of the stellar mass is contained in the core 
and core-torus transition region, which is within about 100~km 
from the rotation axis (see Figure~\ref{fig:mpm}).
The central densities of the hot protoneutron stars are in the sub-nuclear
density regime ($4\times 10^{11}~\rmn{g}~\rmn{cm}^{-3} \la \rho \la 
2\times 10^{14}~\rmn{g}~\rmn{cm}^{-3}$). The structures of these 
protoneutron stars are very different from those of the cold 
neutron stars, which the protoneutron stars will evolve to in roughly 20~s. 
The protoneutron stars have lower central densities, 
rotate less rapidly, and have smaller values of $\beta$.
On the other hand, the structures of 
the three cold neutron stars are similar. Their central densities are 
around $3.5\times 10^{14}~\rmn{g}~\rmn{cm}^{-3}$ and their central 
cores are nearly rigidly rotating with periods of about 1.4~ms, slightly 
less than the fastest observed millisecond pulsar (1.56~ms).

Zwerger and M\"{u}ller \shortcite{zwerger97} performed 2D simulations 
of the core collapse of massive stars. The major difference between their 
models and ours is that they used rather simplified EOS for both 
the pre-collapse and the collapsed models. 
When compared with their fastest rigidly rotating model, AlB3, 
we found their pre-collapse star has less total 
angular momentum and smaller $\beta$ than the pre-collapse white dwarf of 
our Model~I, although both have the same central density. The differences
between their final collapsed models (A1B3G1-A1B3G5) and ours are even 
more significant. The values of $\beta$ of our collapsed objects
are much larger than 
theirs, suggesting that the EOS plays an important role in
the equilibrium configurations of both the pre-collapse white dwarfs and the 
resulting collapsed stars.

The values of $\beta$ of the cold neutron stars are only slightly 
less than the traditional critical value of dynamical instability, 0.27, 
frequently quoted in the literature. The cold neutron stars may still 
be dynamically unstable and a detailed study is required to settle the issue.
Even if they are dynamically stable, they are 
still subject to various kinds of secular instabilities. A rough estimate
suggests that the gravitational-wave driven bar-mode instability dominates. 
The timescale of this instability is about 0.1~s.

\vskip 0.5cm
\noindent {\bf ACKNOWLEDGMENTS}

We thank Kip Thorne and Chris Fryer for stimulating discussions. This research 
was supported 
by NSF grants PHY-9796079, PHY-9900776, and AST-9731698 and NASA grant 
NAG5-4093.


\begin{thebibliography}{}

\bibitem[\protect\citename{Baym, Bethe \& Pethick }%
1971]{baym71} Baym G., Bethe H.\ A., Pethick C.\ J., 1971, Nucl.\ Phys.\ A, 
175, 225

\bibitem[\protect\citename{Bethe \& Johnson }%
1974]{bethe74} Bethe H.\ A., Johnson M.\ B., 1974, Nucl.\ Phys.\ A, 230, 1

\bibitem[\protect\citename{Bisnovatsyi-Kogan \& Blinnikov }%
1974]{bisnovatsyi74} Bisnovatsyi-Kogan G.\ S., Blinnikov S.\ I., 1974, A\&A, 
31, 391

\bibitem[\protect\citename{Bravo \& Garc\'{\i}a-Senz }%
1999]{bravo99} Bravo E., Garc\'{\i}a-Senz D., 1999, MNRAS, 307, 984

\bibitem[\protect\citename{Brown }%
2000]{brown00} Brown J.\ D., 2000, Phys. Rev. D, 62, 084024

\bibitem[\protect\citename{Burrows \& Lattimer }%
1986]{burrows86} Burrows A., Lattimer J.M., 1986, ApJ, 307, 178

\bibitem[\protect\citename{Centrella et al.\ }%
2000]{centrella00} Centrella J.\ M., New K.\ C.\ B., Lowe L.\ L., 
Brown J.\ D., 2000, astro-ph/0010574, submitted to ApJL

\bibitem[\protect\citename{Chandrasekhar }%
1969]{chandrasekhar69} Chandrasekhar S., 1969, Ellipsoidal Figures of 
Equilibrium (New Haven: Yale University Press)

\bibitem[\protect\citename{Cutler \& Lindblom }%
1987]{cutler87} Cutler C., Lindblom L., 1987, ApJ, 314, 234

\bibitem[\protect\citename{Durisen, Gingold \& Tohline }%
1986]{durisen86} Durisen R.\ H., Gingold R.\ A., Tohline J.\ E., 
Boss A.\ P., 1986, ApJ, 305, 281

\bibitem[\protect\citename{Eriguchi \& M\"{u}ller }%
1985]{eriguchi85} Eriguchi Y., M\"{u}ller E., 1985, A\&A, 147, 161

\bibitem[\protect\citename{Finn \& Evans }%
1990]{finn90} Finn L.\ S., Evans C.\ R., 1990, ApJ, 351, 588

\bibitem[\protect\citename{Friedman \& Schutz }%
1975, 1978]{friedman7578} Friedman J.\ L., Schutz B.\ F., 1975, ApJ, 199, L157, 
1978, ApJ, 221, L99

\bibitem[\protect\citename{Goodwin \& Pethick }%
1982]{goodwin82} Goodwin B.\ T., Pethick C.\ J., 1982, ApJ, 253, 816

\bibitem[\protect\citename{Hachisu }%
1986]{hachisu86} Hachisu I., 1986, ApJS, 61, 479

\bibitem[\protect\citename{Hayashi, Eriguchi \& Hashimoto }%
1998]{hayashi98} Hayashi A., Eriguchi Y., Hashimoto M., 1998, ApJ, 492, 286

\bibitem[\protect\citename{van den Horn \& van Weert }%
1981]{horn81} van den Horn L.\ J., van Weert Ch.G., 1981, ApJ, 251, L97

\bibitem[\protect\citename{Houser }%
1998]{houser98} Houser J.\ L., 1998, MNRAS, 209, 1069

\bibitem[\protect\citename{Houser \& Centrella }%
1996]{houser96} Houser J.\ L., Centrella J.\ M., 1996, Phys.\ Rev.\ D, 54, 7278

\bibitem[\protect\citename{Houser, Centrella \& Smith }%
1994]{houser94} Houser J.\ L., Centrella J.\ M., Smith S., 1994, Phys.\ Rev.\
Lett., 72, 1314

\bibitem[\protect\citename{Imamura \& Toman }%
1995]{imamura95} Imamura J.\ N., Toman J., 1995, ApJ, 444, 363

\bibitem[\protect\citename{Imamura, Durisen \& Pickett }%
2000]{imamura00} Imamura J.\ N., Durisen R.\ H., Pickett B.\ K., 2000, ApJ, 
528, 946

\bibitem[\protect\citename{Ipser \& Lindblom }%
1990]{ipser90} Ipser J.\ R., Lindblom L., 1990, ApJ, 355, 226

\bibitem[\protect\citename{Janka \& M\"{o}nchmeyer }%
1989a]{janka89a} Janka H.-Th., M\"{o}nchmeyer R., 1989a, A\&A, 209, L5

\bibitem[\protect\citename{Janka \& M\"{o}nchmeyer }%
1989b]{janka89b} Janka H.-Th., M\"{o}nchmeyer R., 1989b, A\&A, 226, 69

\bibitem[\protect\citename{Lai \& Shapiro }%
1995]{lai95} Lai D., Shapiro W.\ L., 1995, ApJ, 442, 259 

\bibitem[\protect\citename{Lindblom \& Detweiler }%
1977]{lindblom77} Lindblom L., Detweiler S.\ L., 1977, ApJ, 211, 565

\bibitem[\protect\citename{Lindblom \& Detweiler }%
1979]{lindblom79} Lindblom L., Detweiler S.\ L., 1979, ApJ, 232, L101

\bibitem[\protect\citename{Lindblom, Owen \& Morsink }%
1998]{lindblom98} Lindblom L., Owen B.\ J., Morsink S.\ M., 1998, Phys.\ Rev.\ 
Lett., 80, 4843

\bibitem[\protect\citename{Lindblom, Mendell \& Owen }%
1999]{lindblom99}  Lindblom L., Mendell G., Owen B.\ J., 1999, Phys.\ Rev.\ D, 
60, 064006

\bibitem[\protect\citename{M\"{o}nchmeyer \& M\"{u}ller }%
1988]{monchmeyer88} M\"{o}nchmeyer R., M\"{u}ller E., 1988, in NATO ASI 
on {\it Timing Neutron Stars}, ed.\ \"{O}gelman H., D.\ Reidel Publ.\ 
Comp., Dordrecht

\bibitem[\protect\citename{M\"{o}nchmeyer et al.\ }%
1991]{monchmeyer91} M\"{o}nchmeyer R., Sch\"{a}fer G., M\"{u}ller E., 
Katea R.\ E., 1991, A\&A, 246, 417

\bibitem[\protect\citename{M\"{u}ller \& Eriguchi }%
1985]{muller85} M\"{u}ller E., Eriguchi Y., 1985, A\&A, 152, 325

\bibitem[\protect\citename{M\"{u}ller \& Hillebrandt }%
1981]{muller81} M\"{u}ller E., Hillebrandt W., 1981, A\&A, 103, 358

\bibitem[\protect\citename{New, Centrella \& Tohline }%
2000]{new99} New K.\ C.\ B., Centrella J.\ M., Tohline J.\ E., 2000, 
Phys.\ Rev.\ D, 62, 064019

\bibitem[\protect\citename{Nomoto }%
1982]{nomoto82} Nomoto K., 1982, ApJ, 253, 798

\bibitem[\protect\citename{Nomoto }%
1987]{nomoto87} Nomoto K., 1987, in Proc.\ 13th Texas Symposium on Relativistic 
Astrophysics, ed.\ M.\ Ulmer (Singapore: World Scientific)

\bibitem[\protect\citename{Nomoto \& Kondo }%
1991]{nomoto91} Nomoto K., Kondo Y.,1991, ApJ, 367, L19

\bibitem[\protect\citename{Ogata, Iyetomi \& Ichimaru }%
1991]{ogata91} Ogata S., Iyetomi H., Ichimaru S., 1991, ApJ, 372, 259

\bibitem[\protect\citename{Pickett, Durisen \& Davis }%
1996]{pickett96} Pickett B.\ K., Durisen R.\ H., Davis G.\ A., 1996, ApJ, 458, 714

\bibitem[\protect\citename{Rampp, M\"{u}ller \& Ruffert }%
1998]{rampp98} Rampp M., M\"{u}ller E., Ruffert M., 1998, A\&A, 332, 969

\bibitem[\protect\citename{Saijo et al.\ }%
2000]{saijo00} Saijo M., Shibata M., Baumgarte T.\ W., Shapiro S.\ L., 2000, 
ApJ, 548, 919

\bibitem[\protect\citename{Salpeter }%
1961]{salpeter61} Salpeter E.\ E., 1961, ApJ, 134, 669

\bibitem[\protect\citename{Salpeter \& Van Horn }%
1969]{salpeter69} Salpeter E.\ E., Van Horn H.\ M., 1969, ApJ, 155, 183

\bibitem[\protect\citename{Sawyer }%
1989]{sawyer89} Sawyer R.\ F., 1989, Phy.\ Rev.\ D, 1989, 39, 3804

\bibitem[\protect\citename{Shapiro \& Lightman }%
1976]{shapiro76} Shapiro S.\ L., Lightman A.\ P., 1976, ApJ, 207,263

\bibitem[\protect\citename{Shibata, Baumgarte \& Shapiro }%
2000]{shibata00} Shibata M., Baumgarte T.\ W., Shapiro S.\ L., 2000, 
ApJ, 542, 453

\bibitem[\protect\citename{Smith \& Centrella }%
1992]{smith92} Smith S., Centrella J.\ M., 1992, in Approaches to Numerical 
Relativity, ed.\ R.d'Inverno. New York: Cambridge Univ.\ Press

\bibitem[\protect\citename{Smith, Houser \& Centrella }%
1996]{smith96} Smith S., Houser J., Centrella J.\ M., 1996, ApJ, 458, 236

\bibitem[\protect\citename{Stergioulas \& Friedman }%
1998]{stergioulas98} Stergioulas N., Friedman J.\ L, 1998, ApJ, 492, 301

\bibitem[\protect\citename{Strobel, Schaab \& Weigel }%
1999]{strobel99} Strobel K., Schaab C., Wiigel M.\ K., 1999, A\&A, 350, 497

\bibitem[\protect\citename{Tassoul }%
1978]{tassoul78} Tassoul J.-L., 1978, Theory of Rotating Stars, Princeton 
Univ.\ Press, New Jeresy

\bibitem[\protect\citename{Thorne }%
1995]{thorne95} Thorne K.\ S., 1995, in Proc.\ 8th Nishinomiya-Yukawa Memorial 
Symp., on Relativistic Cosmology, ed.\ M.\ Sasaki (Tokyo: Universal Academy 
Press)

\bibitem[\protect\citename{Timmes \& Woosley }%
1992]{timmes92} Timmes F.\ S., Woosley S.\ E., 1992, ApJ, 396, 649

\bibitem[\protect\citename{Tohline }%
1984]{tohline84} Tohline J.\ E., 1984, ApJ, 285, 721

\bibitem[\protect\citename{Tohline, Durisen \& McCollough }%
1985]{tohline85} Tohline J.\ E., Durisen R.\ H., McCollough M., 1985, 
ApJ, 298, 220 

\bibitem[\protect\citename{Williams \& Tohline }%
1988]{williams88} Williams H.\ A., Tohline J.\ E., 1988, ApJ, 334, 449

\bibitem[\protect\citename{Zwerger \& M\"{u}ller }%
1997]{zwerger97} Zwerger T., M\"{u}ller E., 1997, A\&A, 320, 209

\end{thebibliography}
\end{document}